# Resistance of refractory high-entropy alloys to ultrafast laser irradiation

Nikita Medvedev[1,2,*]

*1) Institute of Physics, Czech Academy of Sciences, Na Slovance 1999/2, 182 00 Prague 8, Czech Republic*
*2) Institute of Plasma Physics, Czech Academy of Sciences, Za Slovankou 3, 182 00 Prague 8, Czech Republic*

## Abstract

Response of refractory high-entropy alloys MoNbTaVW and HfNbTaTiZr to ultrafast laser radiation is modelled with the hybrid code XTANT-3, combining tight-binding molecular dynamics with the transport Monte Carlo and Boltzmann equation. A two-temperature state with elevated electronic temperature and a cold atomic lattice is studied. The parameters of the electronic system in such a state are evaluated: electronic heat capacity, thermal conductivity, and electron-phonon coupling parameter with the electronic temperatures up to ~25,000 K. It is also demonstrated that the two refractory alloys do not show signs of nonthermal melting up to the deposited doses of ~10 eV/atom, making them more radiation resistant than the Cantor alloy or stainless steel. These results suggest that heavy-element high-entropy alloys are more radiation resistant than those containing only lighter elements. Damage in irradiated HfNbTaTiZr starts with the selective diffusion of Ti atoms, forming a transient superionic-like state.

## I. Introduction

High-entropy alloys (HEA) are solid solutions consisting of five or more metallic elements with close to equal proportions (also called multi-principal element alloys) [1,2]. Since the discovery of such alloys and demonstration of their exceptional properties, excelling the standard principle-element alloys, they have acquired multiple applications in industry and research requiring high performance, resistance to corrosion, mechanical stresses, extremely high temperatures, and harsh radiation conditions [3]. The possibility of creating materials simultaneously showing high temperature, chemical, and radiation resistance motivates current research in high entropy alloys [4–6].

Recently produced refractory high-entropy alloys (RHEA) contain refractory metals in various chemical compositions [4–6]. RHEA show outstanding performance with respect to mechanical and thermal properties, sustaining high loads [7–10]. These materials also demonstrate extremely high resistance to chemical degradation, such as oxidation, corrosion, or swelling [11].

[*] Corresponding author: email: nikita.medvedev@fzu.cz, ORCID: 0000-0003-0491-1090

They are actively researched for applications in harsh environments, such as aerospace, outer space, nuclear power plants, and plasma diagnostics [6,12–14].

It was reported that W-based RHEA exhibit high radiation resistance under conventional irradiation conditions (low flux or dose-rate), as demonstrated both by theoretical and experimental research [15,16]. Other works investigated RHEA resistance to ion irradiation [17–21]. Their response to ultrafast irradiation (extremely high dose-rate) is still an open question, despite recent growth in their use [13]. For example, RHEA nanoparticles are gaining new applications in targeted photothermal therapy of cancer tissues [22]. Such nanoparticles are produced by femtosecond laser ablation [23].

Two examples of the RHEA are studied in the present work: equiatomic MoNbTaVW and HfNbTaTiZr. They are modelled with the state-of-the-art combined code, XTANT-3, capable of describing simultaneously effects of nonequilibrium, nonthermal modification and evolution of the electronic structure and interatomic potential, and nonadiabatic coupling between electrons and ions (electron-phonon coupling) [24]. The code enables modelling of all the essential effects triggered by ultrafast irradiation, in a sufficiently large simulation box to describe the evolution of the atomic system under ultrafast energy deposition [24,25].

XTANT-3 is used here to evaluate the electronic properties of the two alloys under conditions of electronic excitation as present in femtosecond laser irradiation scenarios. It enables us to obtain the electronic heat capacity, electronic heat conductivity, and the electron-phonon coupling strength up to the electronic temperatures of ~25,000 K [26,27]. Dynamical simulations also provide insights into the RHEA response to irradiation, revealing that MoNbTaVW and HfNbTaTiZr do not exhibit nonthermal melting effects (chemical bonds breaking induced by electronic excitation) up to very high deposited doses, supporting the conclusion on their extreme radiation resistance, even in comparison with the Cantor alloy and stainless steel [28,29].

## II. Model

In this work, the XTANT-3 hybrid simulation tool is applied to study material parameters evolution in the conditions of ultrafast laser irradiation [24]. The code combines a few approaches to comprehensively model the diverse effects of material irradiation: (i) the transport Monte Carlo simulation is used to describe the photoabsorption, electron cascades, Auger decays of core holes (if any produced), and scattering of fast electrons with the atoms; (ii) the Boltzmann collision integrals are applied to trace the slow fraction of electrons populating the conduction band; (iii) the transferable tight-binding method for evaluation of the transient and evolving electronic structure and interatomic forces; (iv) the molecular dynamics simulation for propagation of the atomic trajectories. All the details of the code may be found in its manual [24]. Below, we briefly outline the physical picture and the numerical details used in the code.

The photoabsorption, the induced nonequilibrium electron kinetics, and the Auger decays of produced holes in core atomic orbitals are modelled with the event-by-event individual particle Monte Carlo (MC) approach [30]. The partial photoabsorption cross sections for each atomic shell, the ionisation potentials of core shells, and their Auger-decay times are taken from the EPICS2025 database [31] (see Appendix). The created electrons are then traced until they lose their energy below a chosen cut-off energy of 10 eV, counted from the Fermi energy in the case of metals. The electron impact ionisation cross section is calculated in the framework of the linear response theory (the complex-dielectric function formalism) with the single-pole method [32] (the calculated electron inelastic mean free paths are also shown in the Appendix). This formalism delivers partial differential and total cross sections of scattering on each atomic shell of each element, as well as on the collective conduction band, including plasmon scattering [33]. The quasi-elastic scattering on the atomic system is described with the screened Rutherford cross section with the modified Molier screening parameter, transferring kinetic energy from fast electrons to the atomic ensemble [34].

Slow electrons (with energies below the cutoff) are modelled with the Boltzmann equation, including the electron-electron and electron-phonon interactions [35]:

$$\frac{df_e(\varepsilon_i,t)}{dt} = I_{e-e} + I_{e-a} + I_{MC}. \tag{1}$$

Where $f_e$ is the fractional electronic distribution function defined on the transient energy levels (molecular orbitals obtained with the tight binding formalism, see below); $I_{e-e}$ is the electron-electron scattering integral; $I_{e-a}$ is the electron-atom (or electron-phonon) scattering integral; and $I_{MC}$ is the source term or particles and energy calculated with the MC module above [35].

The electron distribution function relaxes towards the Fermi-Dirac distribution due to the electron-electron interaction. Here, it is calculated with the relaxation time approximation, with the given characteristic time [35]. Selecting the relaxation time to be equal to zero reduces the simulation to instantaneous electron thermalization with the given temperature defined by the deposited energy of the laser pulse, as required further for evaluation of electron-temperature-dependent properties. Electronic nonequilibrium effects are beyond the scope of the present work, since the multitude of those effects requires its own dedicated study [35].

The electron-phonon (electron-ion) coupling is described in the nonperturbative dynamical coupling approach [26], defining the scattering integral as the overlap of the electronic wavefunctions induced by the change of the atomic positions (atomic motion, including vibrations – phonons):

$$I_{e-a}^{ij} = w_{ij}\begin{cases} f(\varepsilon_j)[2-f(\varepsilon_i)] - f(\varepsilon_i)[2-f(\varepsilon_j)]g_{at}(\varepsilon_{ij},T_a), \text{for i} > j \\ f(\varepsilon_j)[2-f(\varepsilon_i)]g_{at}(\varepsilon_{ij},T_a) - f(\varepsilon_i)[2-f(\varepsilon_j)], \text{otherwise} \end{cases}$$
$$w_{ij} \approx \frac{4e}{\hbar\delta t^2}\sum_{\alpha,\beta}\left|c_{i,\alpha}(t)c_{j,\beta}(t_0)S_{i,j}\right|^2 \tag{2}$$

Here, $\varepsilon_i$ are the electronic energy levels (eigenstates of the transient Hamiltonian); for evaluation of the electron scattering probability between the levels *i* and *j*, $w_{ij}$, the wave functions are calculated on two consecutive steps ($t_0$ and $t=t_0+\delta t$) using the linear combination of atomic orbitals (LCAO) basis set within the tight binding Hamiltonian, $\psi_i = \sum_\alpha c_{i,\alpha}\varphi_\alpha$, and $S_{\alpha,\beta}$ is the overlap matrix; *e* is the electron charge, and ħ is the Planck's constant [26]. The electron-phonon coupling parameter is then calculated as follows:

$$G(T_e,T_a) = \frac{1}{V_0(T_e - T_a)}\sum_{i,j}\varepsilon_j I_{e-a}^{ij} \tag{3}$$

Where $V_0$ is the simulation box volume; $T_e$ and $T_a$ are the electronic and atomic temperature, respectively.

The evolution of the electronic energy levels (band structure) is traced with the transferable tight-binding (TB) method [25,36,37]. It includes pairwise interaction of all elements, allowing for modelling of complex materials, such as the RHEA. The diagonalisation of the electronic Hamiltonian (solution of the secular equation *via* the Löwdin orthogonalization method [38]) produces the electronic energy levels (molecular orbitals) and the transient interatomic forces [36] *via* the derivative of the potential energy of the system:

$$U(\{R_{ij}(t)\},t) = E_{rep}(\{R_{ij}(t)\}) + \sum_i f_e(\varepsilon_i,t)\varepsilon_i, \tag{4}$$

where the potential *U* depends on the distances between all the atoms in the simulation box $\{R_{ij}(t)\}$ and the transient electron distribution function; $E_{rep}$ is effective ion-ion repulsion term (containing all contributions apart from the electronic band energies, [36]). The $sp^3d^5$-based PTBP density-functional tight-binding parametrisation is employed to construct the Hamiltonian and the overlap matrix [39].

As can be seen from Eq.(4), any change in the electronic distribution function directly affects the interatomic potential. Thus, the method employed is capable of describing the nonthermal melting: disruption of the atomic bonds due to electronic excitation, even without nonadiabatic atomic heating (electron-phonon coupling) [25,40]. Nonthermal melting was theoretically shown to take place in stainless steel and Cantor alloy, and, therefore, is an important mechanism to study in other high entropy alloys [28,29].

The molecular dynamic simulations trace the evolution of the atomic coordinates, in response to both changes in the interatomic potential (nonthermal effects) and electron-phonon energy exchange (thermal heating) [25]. In the current simulation, Martyna-Tuckerman’s 4th-order algorithm is employed [41]. We use the time-step of 0.5 fs for the dynamical simulations, and 0.1 fs in the evaluation of the electron-phonon coupling parameter. The energy transfer from electron scattering is delivered to atoms via the velocity scaling algorithm on each simulation time step.

The electron heat capacity is obtained as the derivative of the electronic entropy at the given electron temperature [42]:

$$C_e(T_e,T_a) = \frac{1}{V_0}\sum_i \frac{\partial f_e(\varepsilon_i)}{\partial T_e}(\varepsilon_i - \mu(T_e)), \tag{5}$$

Where $\mu(T_e)$ is the electron chemical potential. The derivative of the distribution function is evaluated numerically on the given set of energy levels in the TB module.

The electron heat conductivity is calculated in the framework of the Onsager coefficients, including the electron-phonon and electron-electron contributions *via* the Matthiessen rule [27]:

$$\begin{gathered} \kappa_{tot} = \left(\frac{1}{\kappa_{e-a}} + \frac{1}{\kappa_{e-e}}\right)^{-1} \\ \kappa_{e-a} = L_{22} - T_e\frac{L_{12}^2}{L_{11}} \\ L_{ij} = -\frac{(-1)^{i+j}}{V_0 m_e}\sum \frac{df}{d\varepsilon_k}(\varepsilon_k - \mu)^{i+j+2}\,|\langle k|p|k'\rangle|^2 \end{gathered} \tag{6}$$

Where $m_e$ is the free electron mass; and the oscillator strength coefficients $|\langle k|p|k'\rangle|^2$ are calculated with the tight-binding formalism [27]. Electronic contribution is evaluated analogously, with its own scattering matrix element [27].

Both, calculation of the electron heat capacity and conductivity are performed on the Monkhorst-Pack *k*-vector grid of 7x7x7 points in the supercell [27,43].

MoNbTaVW is modelled with 250 atoms in the simulation box placed in a BCC structure; HfNbTaTiZr is modelled with 320 atoms in an FCC structure. The number of atoms is defined by the constraints of the equiatomic stoichiometry of the alloys, and the considered structure (2 atoms in the unit cell of BCC structure, 4 atoms in the FCC unit cell), and are chosen to be sufficiently large to eliminate finite-size effects, as discussed in [26].

The atoms are placed randomly on the corresponding crystalline grid. The steepest descent algorithm is used to minimize the potential energy, adjusting the materials density. Then, the simulation box is allowed to thermalise at room temperature for a few hundred femtoseconds prior to the simulation of irradiation. The simulation box sizes are 15.5 x 15.5 x 15.5 $Å^3$ for MoNbTaVW and 16.58 x 16.56 x 20.72 $Å^3$ for HfNbTaTiZr. That corresponds to the atomic density of 13.48 $g/cm^3$ (MoNbTaVW) and 11.05 $g/cm^3$ (HfNbTaTiZr), in a reasonable agreement with the experimental values of around 12.56 $g/cm^3$ and 9.9-10.0 $g/cm^3$, respectively [44–46]. Some deviations of the densities from the experimental values are expected for multi-element transferrable DFTB simulations, since the tight binding parameters are not specifically fitted for particular materials but instead reproduce a wide variety of them.

Initial atomic velocities are set randomly according to the Maxwellian distribution at room temperature. Periodic boundary conditions are used along all the axes, mimicking bulk material, with the microcanonical ensemble. All the illustrations of the atomic snapshots are prepared with the help of OVITO [47]. XTANT-3 was previously validated against various experiments, showing a good agreement in reproducing damage thresholds, transient electronic parameters, and atomic dynamics [25–27].

## III. Results

We start by evaluating the electronic properties of the two RHEA, which are the essential parameters in standard modelling of laser-irradiation of metals, such as the two-temperature model or two-temperature molecular dynamics [48,49].

The partial electronic densities of states in MoNbTaVW and HfNbTaTiZr are shown in the Appendix. Having the density of states and the numerically evaluated chemical potential, the specific electronic heat capacities may be calculated at increased electronic temperatures, see Figure 1. As is typical for metals [50], it first rises nearly linearly, reaching a peak at ~10,000-12,000 K, decreasing afterwards. The peak reached in MoNbTaVW is higher than the one in HfNbTaTiZr, which is consistent with the fact that the number of electrons in the conduction band of MoNbTaVW is 5.4 per atom vs. 4.4 electrons per atom in HfNbTaTiZr.

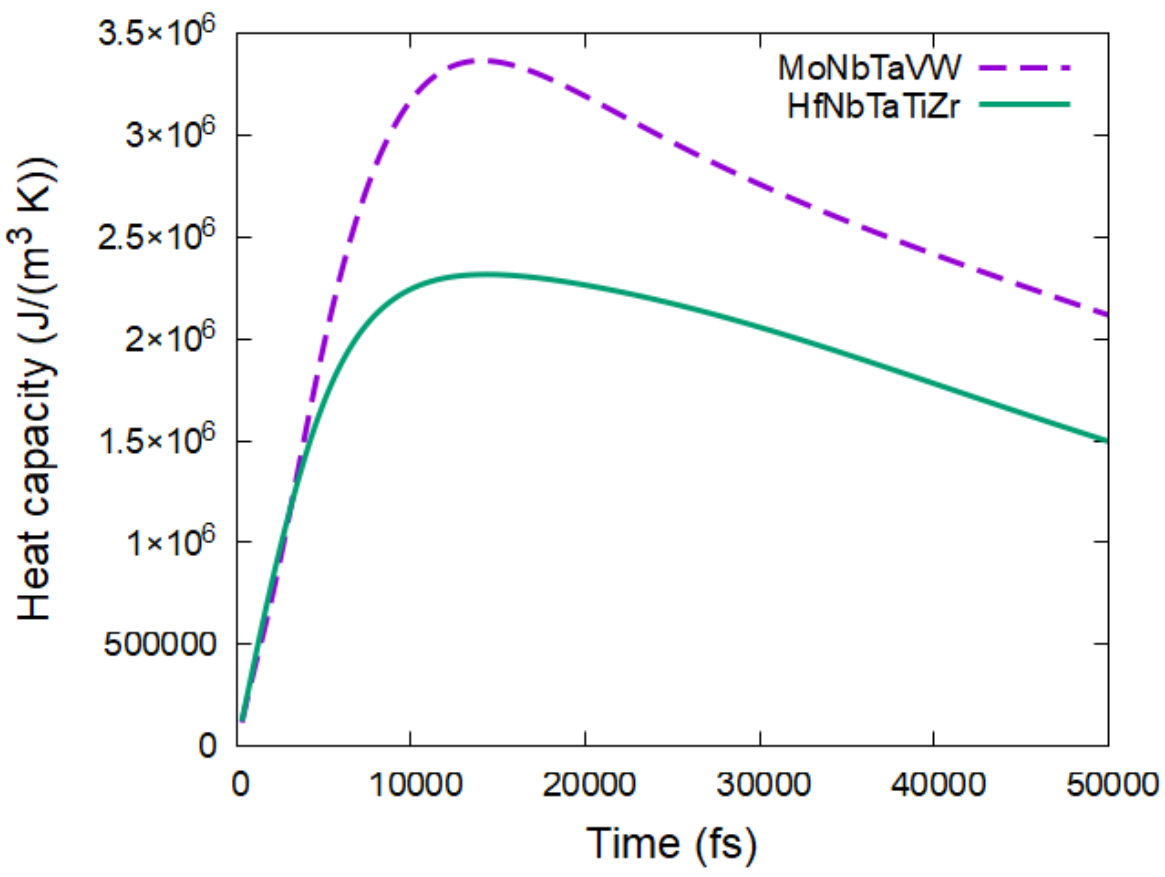

*Figure 1. Electronic heat capacity in MoNbTaVW and HfNbTaTiZr calculated with XTANT-3.*

The electron heat conductivities are shown in Figure 2. Here, also, the peak is higher in MoNbTaVW than in HfNbTaTiZr, creating larger electronic heat conductivity at elevated electronic temperatures. Both are noticeably higher than the electronic heat conductivity in the Cantor alloy and stainless steel [28,29]. This suggests faster heat dissipation in the RHEA, making them more resistant to electron excitation, such as laser irradiation.

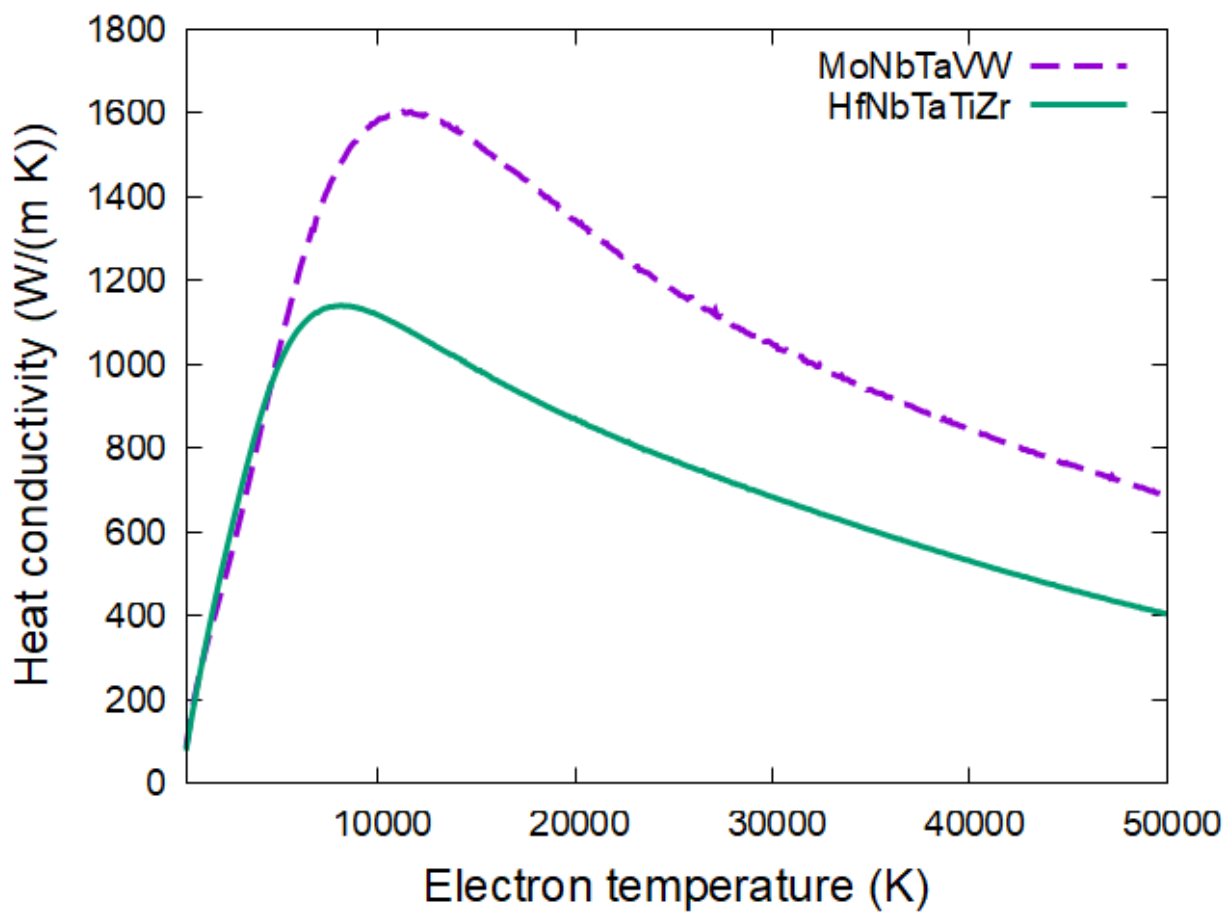

*Figure 2. Electronic heat conductivity in MoNbTaVW and HfNbTaTiZr calculated with XTANT-3.*

The calculated electron-phonon (electron-ion) coupling is shown in Figure 3. The peak values are, again, higher in MoNbTaVW than in HfNbTaTiZr, but both are lower than the coupling strength in the Cantor alloy and stainless steel [28,29]. Slower atomic heating by electrons also points to higher resistance to ultrafast laser irradiation, allowing the heat to dissipate in the electronic system before being transferred to the atomic one.

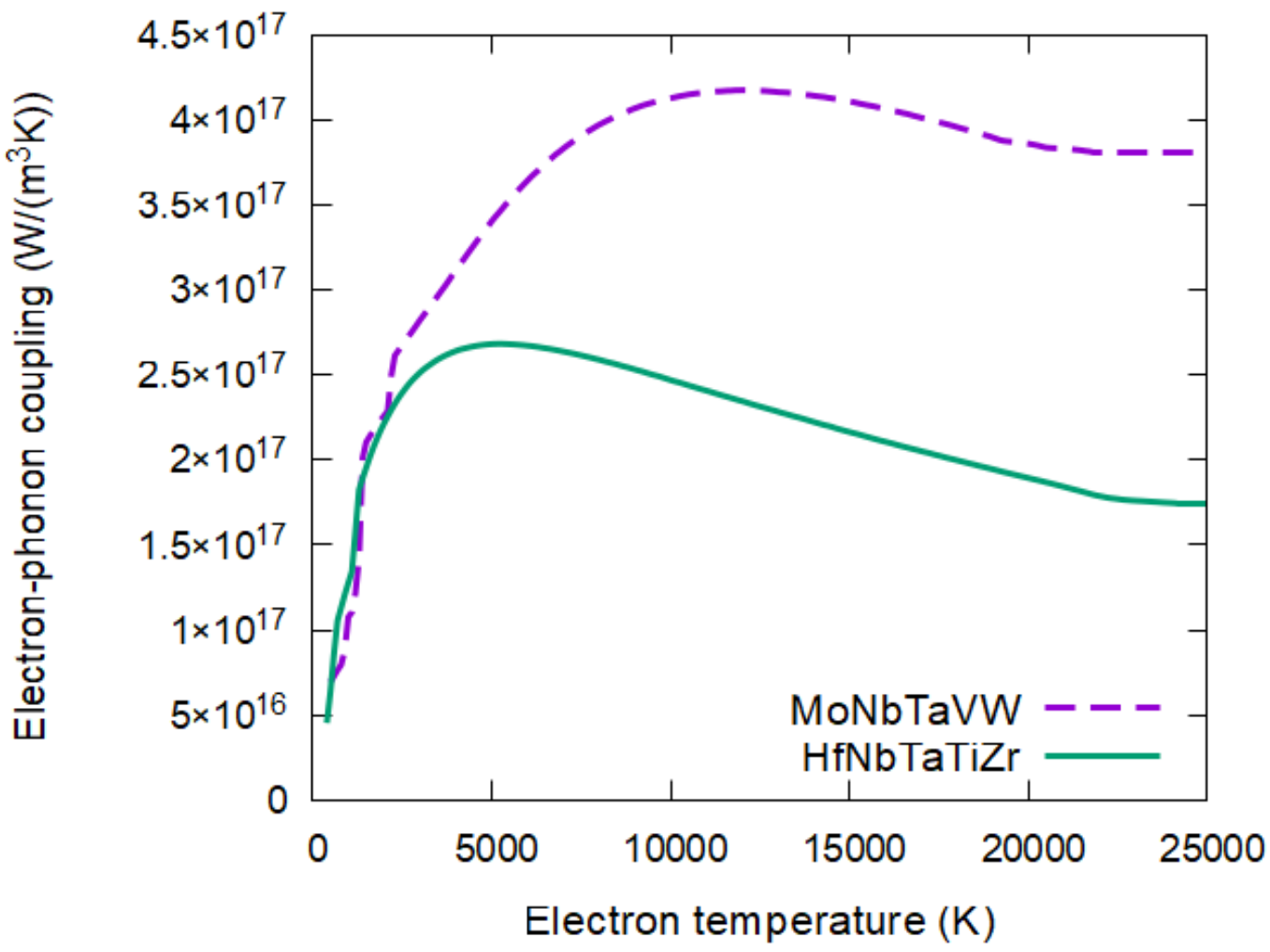

*Figure 3. Electron-phonon (electron-ion) coupling parameter in MoNbTaVW and HfNbTaTiZr calculated with XTANT-3.*

In a series of dynamic simulations within the Born-Oppenheimer-like simulations (artificially excluding the electron-phonon coupling [35]), neither of the studied RHEA showed signs of damage up to the deposited doses of 10 eV/atom (corresponding to the electronic temperatures of $T_e$~40,000 K in MoNbTaVW and Te~46,000 K in HfNbTaTiZr). Electronic excitation does not induce nonthermal melting in the bulk of these materials (surface effects and nonthermal melting due to expansion of nano-sized materials are generally expected in metals, but are beyond the scope of the current work [51]). This suggests that neither BCC nor FCC refractive high-entropy alloys are susceptible to nonthermal melting.

Including the electron-phonon coupling in the simulations leads to energy exchange between the electronic and the atomic systems, increasing the atomic temperature, which may trigger thermal melting. In MoNbTaVW, at doses of ~5 eV/atom (peak $T_e$~21,000 K), the melting onsets within 5 ps, starting from the displacements of vanadium atoms, closely followed by all the others except tungsten, which requires longer times. At lower doses, the melting did not occur within the simulation time of 5 ps, but is expected to take place at longer times for the doses above ~0.35 eV/atom (corresponding to the melting temperature in MoNbTaVW of ~2760 K [52]).

Simulations of HfNbTaTiZr with the electron-phonon coupling included show that at a dose above ~1.1 eV/atom, first damage occurs within 5 ps simulation time, see example of atomic snapshots in Figure 4. Melting at longer timescales is expected to take place at the doses above ~0.33 eV/atom, producing atomic temperatures above the melting point in HfNbTaTiZr [53].

Interestingly, the damage in HfNbTaTiZr at the doses above ~1.1 eV/atom takes place only in the titanium subsystem (within 5 ps), whereas all other sublattices are intact, indicating a transient radiation-induced superionic state or selective disordering: liquid-like Ti within solid other lattices. This can be seen in the mean displacements of Ti atoms, shown in Figure 5, indicating that Ti atoms start diffusing out of their equilibrium positions, whereas displacements of all other atoms are saturated at this timescale.

This behaviour is not accompanied by a selective increase in the atomic temperatures – all elements have temperatures close to the average atomic one (top panel in Figure 5), in contrast to the effects observed in the Cantor alloy and stainless steel [28,29]. Also note that the superionic state occurs already after equilibration between the electronic and atomic temperatures, indicating its thermal, rather than nonthermal [54], nature.

At the dose of ~3 eV/atom, Nb atoms start to displace within 5 ps, and with the further increase in the dose, all the atoms follow, completely disordering on a few-picosecond timescale.

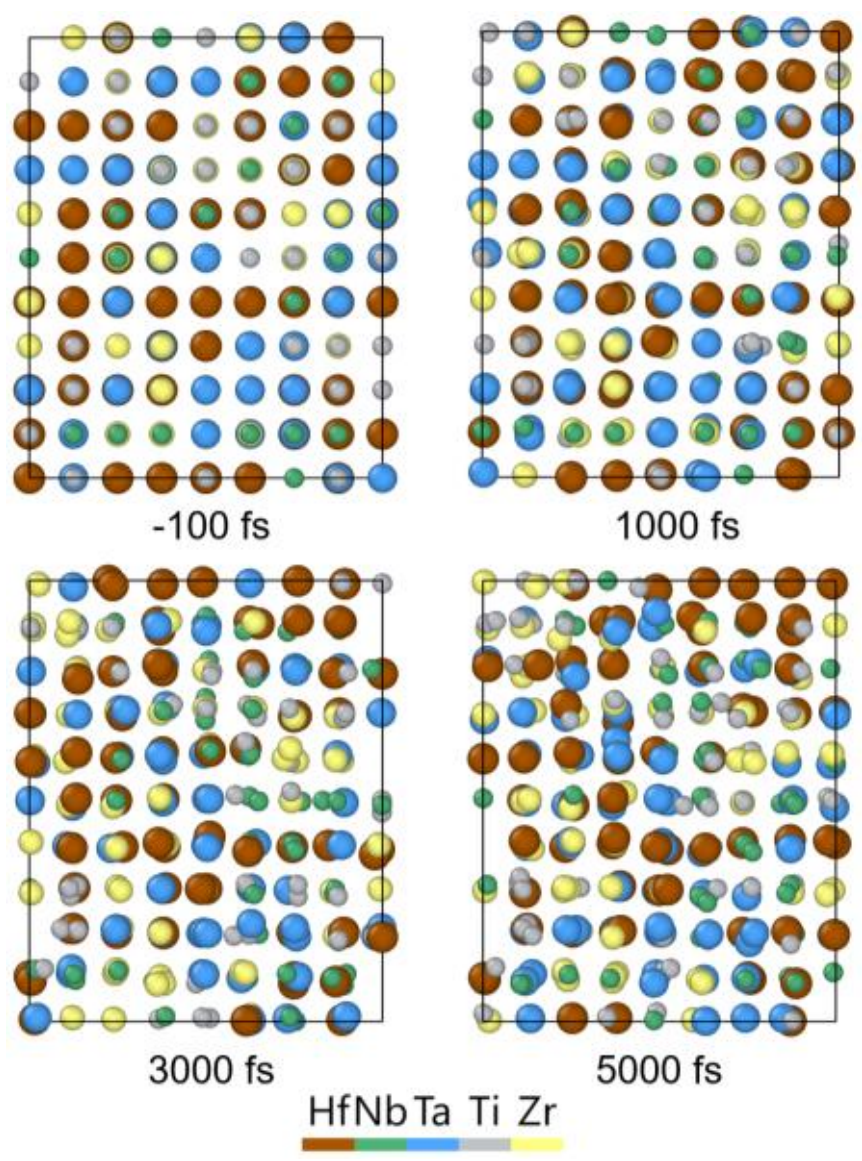

*Figure 4. Atomic snapshots of HfNbTaTiZr irradiated with a 10 fs FWHM pulse, 92 eV photon energy, 2 eV/atom deposited dose, calculated with XTANT-3.*

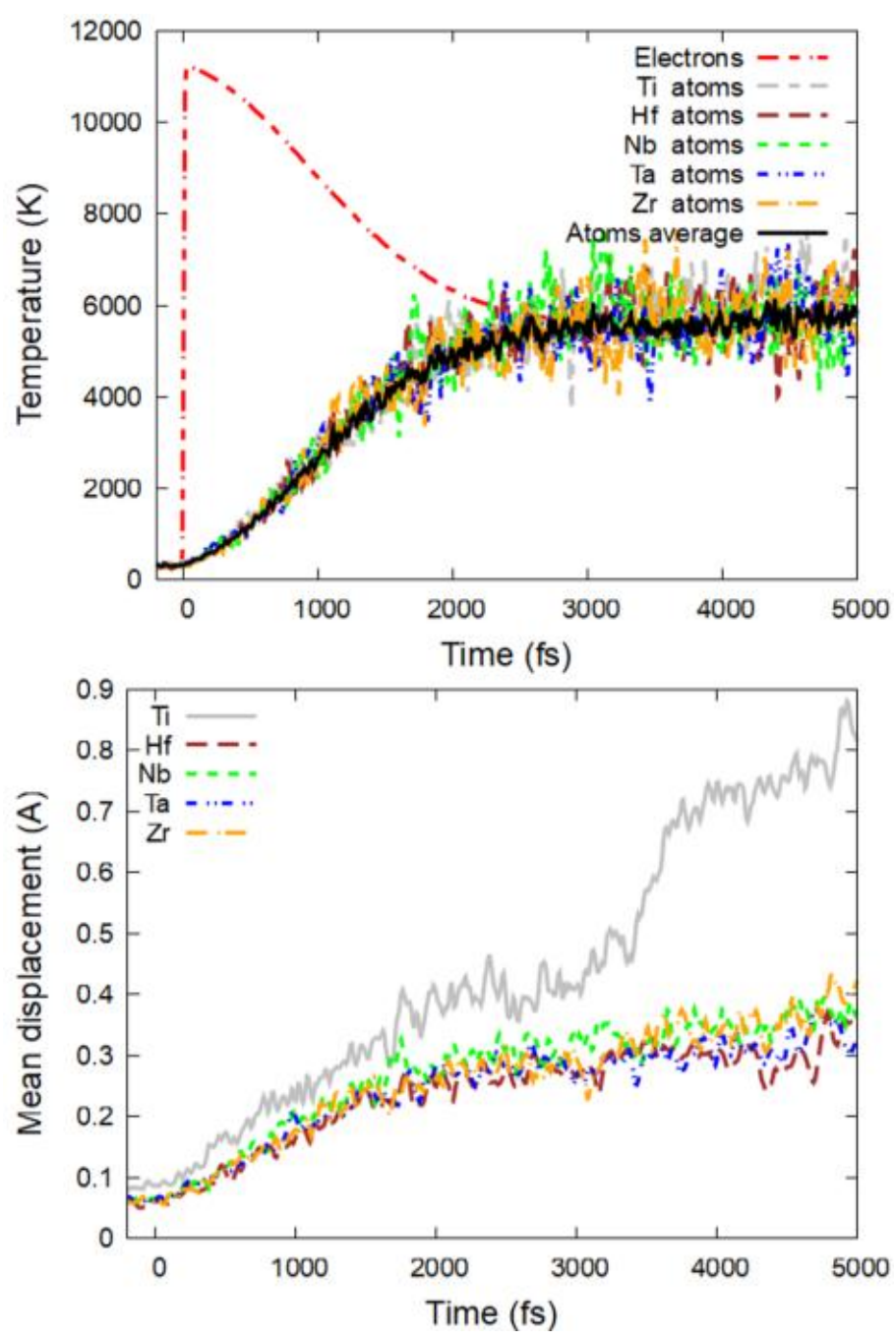

*Figure 5. (Top panel) Electronic, total, and element-specific atomic temperatures; (bottom panel) element-specific mean displacements in HfNbTaTiZr irradiated with a 10 fs FWHM pulse, 92 eV photon energy, 2 eV/atom deposited dose, calculated with XTANT-3.*

The damage threshold doses may be converted into the incident fluence, assuming normal incidence and no transport effects: $F = D\lambda n_{at}$, with $D$ being the threshold dose, λ the photon attenuation length (see Appendix), and $n_{at}$ the atomic concentration. The corresponding damage threshold fluences for both RHEA are shown in Figure 6. The thermal melting thresholds are very close in HfNbTaTiZr and MoNbTaVW, owing to close threshold doses and photon attenuation lengths. Kinks and jumps in the curves correspond to the onset of photoabsorption by various atomic core shells (see Appendix). The estimated threshold fluences and doses can be used to guide future experiments and applications involving irradiation of RHEA.

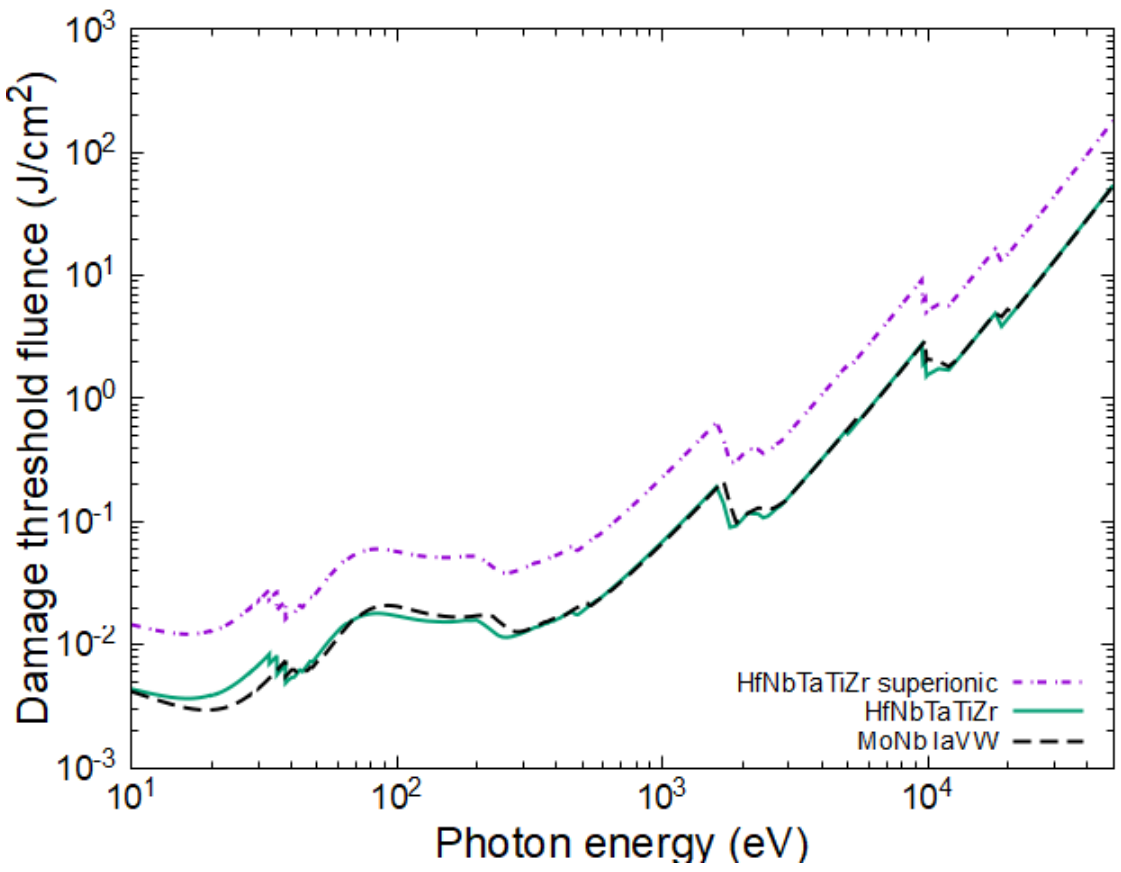

*Figure 6. Estimated damage threshold fluences for thermal melting and ultrafast transient superionic transition in HfNbTaTiZr and melting in MoNbTaVW.*

## IV. Conclusions

Equiatomic refractory high-entropy alloys (RHEA), MoNbTaVW and HfNbTaTiZr, are studied with the XTANT-3 simulation tool, describing electronic excitation, modification of the electronic structure, changes in the interatomic potential induced by electronic excitation, and electron-phonon coupling. Electronic heat capacity, conductivity, and electron-phonon coupling are calculated for electronic temperatures up to ~25,000 K.

It is revealed that, in contrast to the Cantor alloy and stainless steel, neither of the RHEA shows signs of nonthermal melting (a disruption of interatomic bonds directly by electronic excitation) at least up to the deposited dose of 10 eV/atom. Instead, both materials melt thermally, via heating of the atomic system by electron-phonon coupling. Interestingly, at the deposited doses above ~1.1 eV/atom, HfNbTaTiZr exhibits ultrafast transient superionic state – selective melting of Ti sublattice within intact lattices formed by other elements.

The damage threshold fluences in a wide range of photon energies are evaluated for both RHEA, which can be used to guide future experiments and applications.

## V. Conflicts of interest

There are no conflicts to declare.

## VI. Data and code availability

The code XTANT-3 used to obtain electronic properties and simulate irradiation effects, and input data, including photon and electron mean free paths, are available from [24]. The tables with the calculated electronic heat capacity, conductivity, and electron-phonon coupling parameter are available from [55].

## VII. Acknowledgements

Computational resources were provided by the e-INFRA CZ project (ID:90254), supported by the Ministry of Education, Youth and Sports of the Czech Republic. The author thanks the financial support from the Czech Ministry of Education, Youth, and Sports (grant nr. LM2023068), and from the European Commission Horizon MSCA-SE Project MAMBA [HORIZON-MSCA-SE-2022 GAN 101131245].

## VIII. Appendix

Figure 7 and Figure 8 show the calculated electronic densities of states in MoNbTaVW and HfNbTaTiZr, respectively. The DOS in MoNbTaVW is similar to that from [5], whereas in HfNbTaTiZr, it is similar to that calculated in [7].

The photon mean free paths, constructed from the atomic contribution accounting for the stoichiometry of the materials, and the electronic mean free path, calculated with the single-pole complex dielectric function formalism, in MoNbTaVW and HfNbTaTiZr, are shown in Figure 9 and Figure 10, respectively.

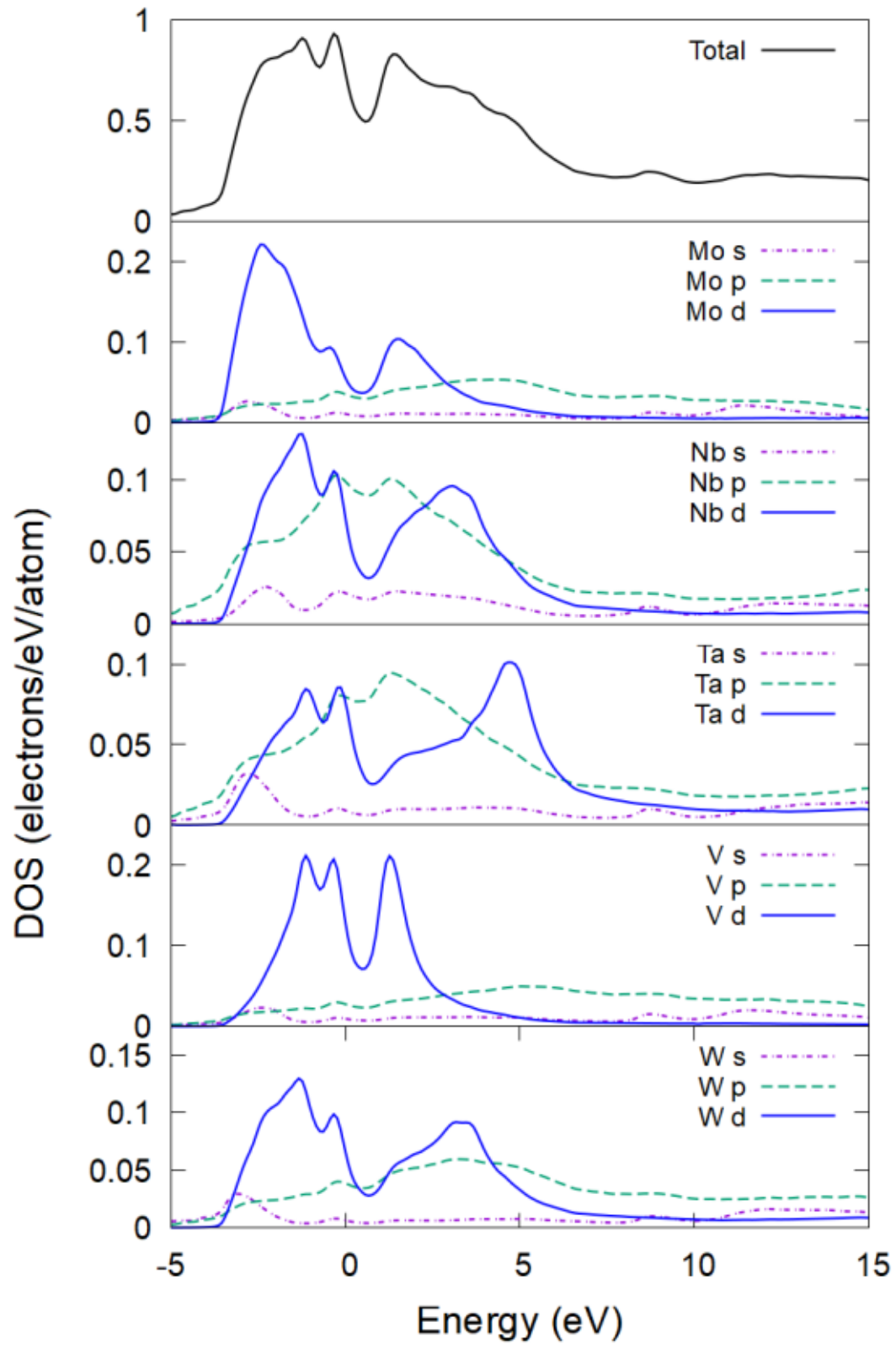

*Figure 7. Total and partial DOS in MoNbTaVW counted from the Fermi level, calculated with XTANT-3.*

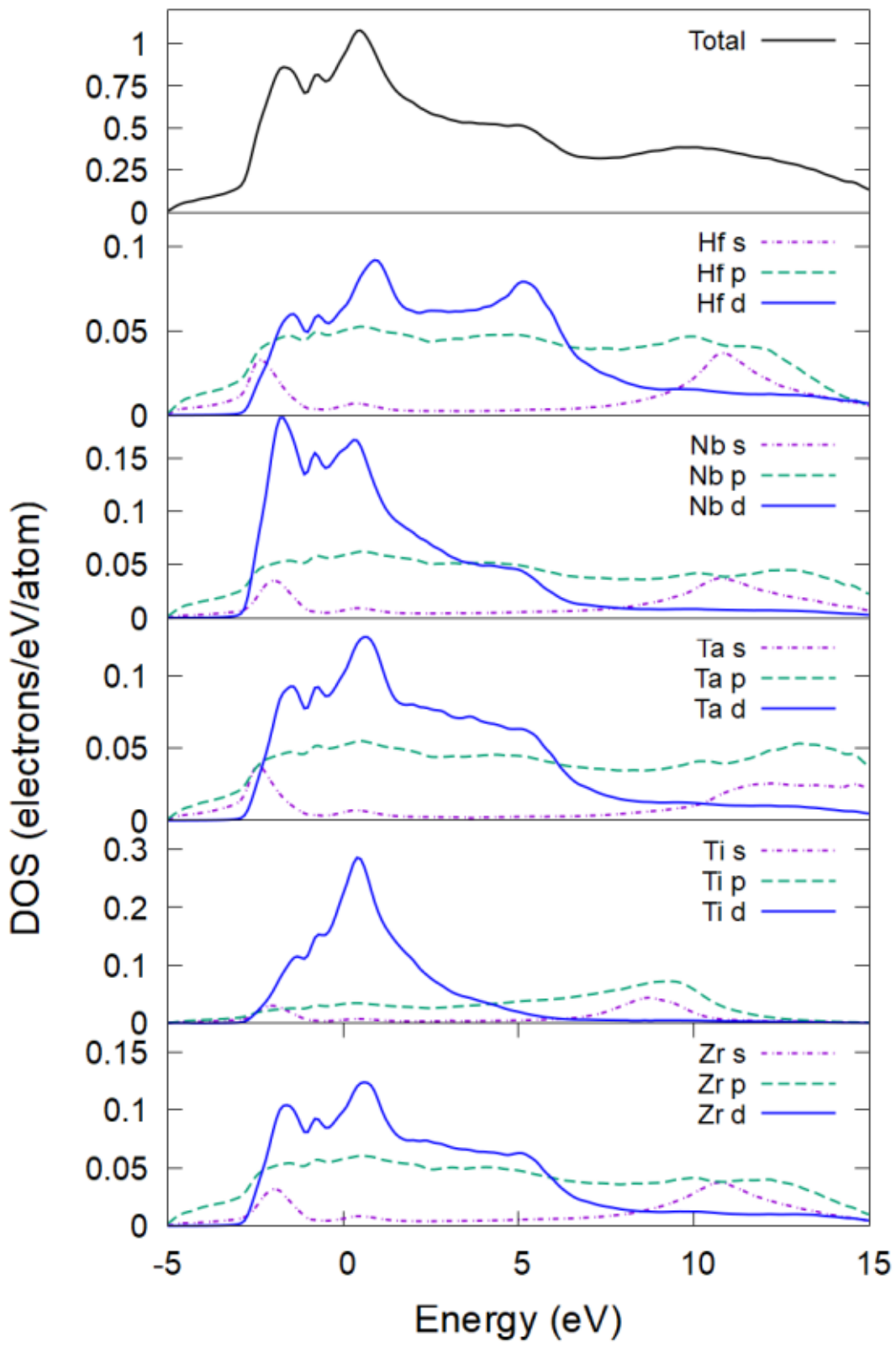

*Figure 8. Total and partial DOS in HfNbTaTiZr counted from the Fermi level, calculated with XTANT-3.*

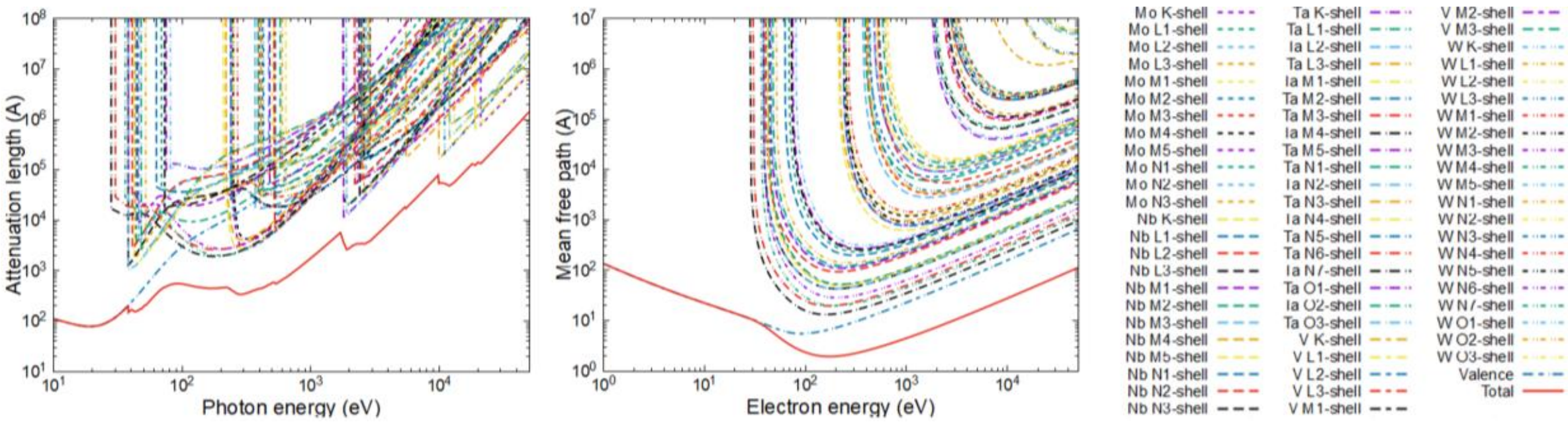

*Figure 9. Element and shell-resolved photon attenuation length and electron mean free paths in HfNbTaTiZr.*

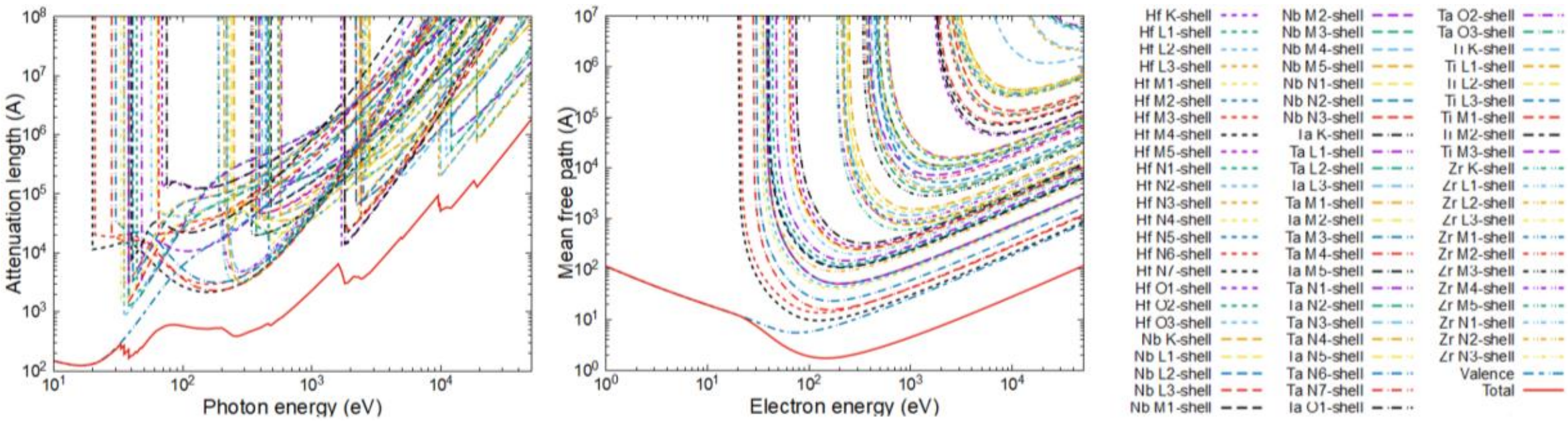

*Figure 10. Element and shell-resolved photon attenuation length and electron mean free paths in HfNbTaTiZr.*